\documentstyle[preprint,aps,epsfig]{revtex}
\tightenlines
 
\begin{document}
\newcommand{\beq}{\begin{equation}}
\newcommand{\eeq}{\end{equation}}
\newcommand{\beqa}{\begin{eqnarray}}
\newcommand{\eeqa}{\end{eqnarray}}
\newcommand{\sr}{\sqrt}
\newcommand{\fr}{\frac}
\newcommand{\mn}{\mu \nu}
\newcommand{\G}{\Gamma}

\draft
\preprint{ INJE-TP-00-04}
\title{Stability of the GRS Model}
\author{ Y.S. Myung\footnote{E-mail address:
ysmyung@physics.inje.ac.kr} and Gungwon Kang\footnote{E-mail address:
kang@physics.inje.ac.kr} }
\address{
Department of Physics, Inje University,
Kimhae 621-749, Korea}
\maketitle
\begin{abstract}

We discuss the compatibility between the weaker energy condition 
and the stability of Gregory, Rubakov and Sibiryakov (GRS) model. 
Because the GRS spacetime violates the weak energy condition, it may 
cause the instability. In the GRS model, the four dimensional gravity 
can be described by the massive KK modes with the resonance. Hence, 
instead of considering the weaker energy condition, we require for 
the stability of this model: no tachyon and no ghost condition for 
graviton modes ($h_{\mu\nu}$). No tachyonic condition ($m^2_h \geq 0$) 
is satisfied  because the lowest state $m_h=0$ is supersymmetric 
vacuum state. Further, no ghost state condition is achieved if one 
requires some relations for the matter source: 
$2T_{55}= T^{\mu}_{\mu}=3(T_{22}+T_{33})$. It turns out that, although
the GRS spacetime does not satisfy the weaker energy condition, it is
stable against small perturbation.  
 
\end{abstract}
\bigskip
 
\newpage
 
\section{Introduction}
 
Recently, there have been lots of interest in the phenomenon
of localization of gravity proposed by Randall and Sundrum
(RS)~\cite{RS} (for previous relevant work see references 
therein). RS assumed a single positive tension 3-brane and 
a negative bulk cosmological constant in the five dimensional 
(5D) spacetime. By considering metric fluctuations from a background
which is isomorphic to sections of 5D anti-de Sitter spacetime 
(${AdS}_5$), they have shown that it reproduces the effect of four 
dimensional (4D) gravity localized on the brane without the need 
to compactify the extra dimension due to the ``warping" in the 
fifth dimensional space. In more detail, the solution to 
linearized equations in the five dimensions results in a zero 
mode, which can be identified with the 4D massless graviton, and 
massive continuum Kaluza-Klein (KK) modes. Surprisingly, 
the wavefunctions of the massive continuum KK modes are suppressed 
at the brane for small energies, and thus ordinary gravity 
localized on the brane is reproduced at large distances.      

Gregory, Rubakov and Sibiryakov (GRS)~\cite{GRS} have recently 
considered a brane model which is not asymptotically $AdS_5$, 
but Minkowski flat. In the GRS model, the ordinary 4D Newtonian 
potential is reproduced from not the massless zero mode, but the 
resonance of zero mass in the continuum KK 
spectrum~\cite{GRS,CEH,DGP,Witten}. In this sense the GRS model of 
``a resonance graviton" would differ from the RS model. However, 
there exist two potential problems with this model: One is the 
mismatch in polarization states and the other is the violation of 
weaker energy condition (WEC). It was pointed out that a massive 
graviton propagator with $3$ polarization states does not reproduce 
the massless graviton propagator with $2$ due to the missmatch of 
the number of polarization states~\cite{DGP}. Contrary to it, 
Cs\'aki, Erlich and Hollowood~\cite{CEH2} have argued that in the 
presence of localized source at $y=0$ the effect ($\xi^5$) of the 
bending of the brane exactly compensates for the extra polarization 
in the massive graviton propagator. Thus the $m^2_h \rightarrow 0$ 
limit of the massive propagator at intermediate scales is equivalent 
to the massless propagator of the Einstein theory just as in the RS 
scenario. At ultra large scales, however, this theory includes scalar 
anti-gravity~\cite{GRS2}, which may be cured by the RG 
analysis~\cite{CEHT}. The most important fact is probably that the 
mechanism to cancel the extra polarization gives arise to the ghost 
problem~\cite{DGP2}. In fact, the role of the radion field with a 
negative kinetic term is discussed in models with metastable 
graviton~\cite{PRZ}. In order to have a well-defined theory, 
the ghost should disappear.   

We have introduced the trace field ($h$) in the RS model~\cite{KM} 
instead of $\xi^5$ in Ref.~\cite{GT}. Instead of the localized 
source ($T_{\mu\nu}(x,y) =T_{\mu\nu}(x)\delta (y)$), we introduce 
a matter source with uniform trace along the extra dimension 
($T^{\mu}_{\mu}(x,y) =T^{\mu}_{\mu}(x)$). 
Fortunately, it is shown that massive graviton 
modes contain ghost states which can be removed by assuming a further 
condition on the matter source. 

The WEC is a basic requirement. In the GRS model we find that 
this is violated. But the compatibility between the WEC and the
recovery of Einstein gravity seems to be not so important. In the
brane world, the first thing that we have to do is to recover the
Einstein gravity. A more important thing is that the weak energy 
condition may be closely related to the stability of the GRS 
spacetime. However, the actual stability analysis of a nonlinear 
system of the GRS model means to assess the reliability of its 
linearized approximation~\cite{Bill}. 

In this paper we wish to study the stability of the GRS background. 
We require that the stability of this spacetime be given by two:  
\begin{enumerate}
\item There are no tachyons for $h_{\mu\nu}(x)$.
\item $h_{\mu\nu}(x)$ has no ghost (i.e., no negative norm state).
\end{enumerate}
We will show that although the GRS spacetime does not satisfy 
the WEC, this spacetime is stable. In this paper,
we use the signature $(-, +, +, +, +)$ and MTW conventions.

\section{Weaker energy condition (WEC)}

In this section we explicitly show that the GRS model does not satisfy
the weakest form of a positive energy condition, which is the so-called
null energy condition or the weaker energy condition, saying that 
the stress-energy tensor $T_{MN}$ obeys $T_{MN}\xi^M\xi^N 
\geq 0$ for any null vector $\xi^M$. As pointed out by 
Witten~\cite{Witten}, given such energy condition, a ``holographic 
$c$-theorem" for the ${AdS}_5$ says that as one approaches 
to spatial infinity in the extra dimension, the bulk cosmological 
constant $\Lambda$ can only become
more negative. The bulk cosmological constant in the GRS model is a 
negative constant in the vicinity of the positive tension brane at the
center, but vanishes beyond the negative tension branes. 

Let us consider a five-dimensional spacetime which is described by the
metric as 
\beq
ds^2 = \hat{g}_{MN} dx^Mdx^N
= e^{2A(y)} g_{\mu\nu}dx^{\mu}dx^{\nu} +dy^2,
\eeq
where $g_{\mu\nu}$ is Ricci flat (i.e., $R_{\mu\nu}(g)=0$) and we 
assume that the background matter ($T^{(0)}_{MN}$) producing such metric
is distributed only on four-dimensional domain walls 
(i.e., $T^{(0)}_{yy}=0$). Then the non-vanishing components of the Ricci
tensor are~\cite{FGRW}
\beq
\hat{R}_{\mu\nu} = - \big [ A'' +4(A')^2 \big ] \hat{g}_{\mu\nu}, 
\qquad\qquad \hat{R}_{yy} =-4 \big [ A'' +(A')^2 \big ] .
\eeq
Here $A'=\partial_y A$. Note that the contribution to the background 
matter stress tensor coming from the cosmological constant does not 
affect the WEC because it is proportional to the metric. Hence we
separate $\Lambda$ from $T^{(0)}_{MN}$. Using the Einstein equation 
$\hat{G}_{MN} =\hat{R}_{MN} -\fr{1}{2}\hat{R}\hat{g}_{MN} 
= -\Lambda \hat{g}_{MN} +8\pi G_5 T^{(0)}_{MN}$, 
we see for any null vector $\xi^M$
\beqa
T^{(0)}_{MN}\xi^M\xi^N &=& \fr{1}{8\pi G_5} \hat{R}_{MN}\xi^M\xi^N 
\nonumber  \\
&=& \fr{1}{8\pi G_5} \Big \{ - \big [ A'' +4(A')^2 \big ]
\xi_{\mu}\xi^{\mu} -4 \big [ A'' +(A')^2 \big ] (\xi^y)^2 
\Big \}  \nonumber \\
&=& -\fr{3}{8\pi G_5} A'' (\xi^y)^2.
\eeqa
On the last line we used $\xi_M\xi^M=\xi_{\mu}\xi^{\mu} 
+(\xi^y)^2 =0$. 

Therefore, the WEC for the background matter $T^{(0)}_{MN}\xi^M\xi^N
\geq 0$ for any null vector $\xi^M$ becomes equivalent to 
$A'' \leq 0$. Subsequently, $A'$ must be a non-increasing monotonic 
function in the coordinate $y$. From the $yy$-component of the 
Einstein equation, we also find 
\beq 
\Lambda (y) = -6(A')^2,
\eeq
which implies that as one goes to large $y$, the cosmological constant
can only become more negative provided that $A'$ was a negative value
at some point.

\begin{figure}[tbp]
%\epsfysize=4.5cm
%\hspace{2.5cm}
\begin{center}
\epsfig{file=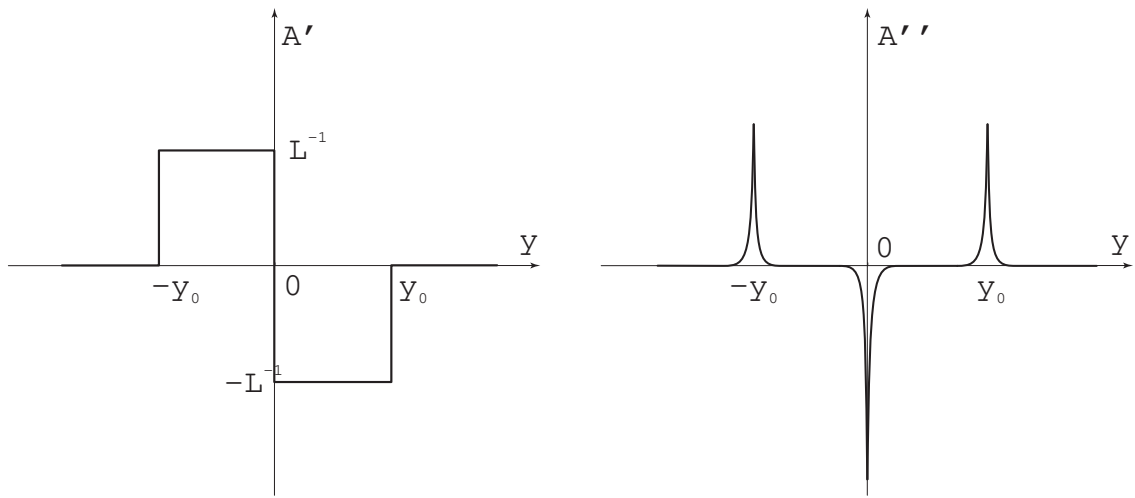,width=15cm,height=6.5cm}
\vspace{1cm}
\caption{The behavior of $A'$ and $A''$}
\label{fig}
\end{center}
\end{figure}

For the perfect ${AdS}_5$ spacetime which corresponds to $\Lambda 
={\rm constant} < 0$ and $T^{(0)}_{MN} =0$, $A_{AdS}(y) = -y/L$ 
where $L$ is the radius of ${AdS}_5$. Thus the perfect anti-de Sitter 
spacetime leads to $A' =-1/L$, $A''=0$, and so there is no RG flow 
at all. For the RS spacetime, $A_{RS}(y) = -|y|/L$. Then $A'_{RS} (y) 
= -\theta (y)/L$, $A''_{RS}(y)= -2\delta(y)/L \leq 0$, and so the RS
model satisfies the WEC. For the GRS spacetime, however, 
\beq
A_{GRS} (y) =
\left\{
\begin{array}{cc}
-|y|/L \qquad\qquad {\rm for} \qquad |y| \leq y_0, \\
-y_0/L \qquad\qquad {\rm for} \qquad |y| \geq y_0.
\end{array}\right.
\eeq
Therefore, we have
\beq
A'_{GRS} (y) =
\left\{
\begin{array}{cc}
-\theta (y)/L \qquad {\rm for} \qquad
|y| \leq y_0,  \\
0 \qquad\qquad {\rm for} \qquad  |y| \geq y_0,
\end{array}\right.
\qquad\qquad 
A''_{GRS} (y) = -\fr{2}{L} \delta^{GRS} (y), 
\eeq
where $\delta^{GRS} (y) \equiv \delta (y)
-\fr{1}{2}\delta (y-y_0) -\fr{1}{2} \delta (y+y_0)$. 

As is shown in Fig.~\ref{fig}, $A'_{GRS} (y)$ is not a monotonically
decreasing function and thus the condition of $A'' \leq 0$ does not 
hold. On the position of the negative branes 
($y=\pm y_0$), one finds $A'' (y) = \delta (y\pm y_0)/L$. 
Of course, this indicates the violation of $A'' (y) \leq 0$. 
We wish to emphasize that $y=\pm y_0$ are just points at which the 
negative branes are introduced to obtain the Minkowski spacetime 
for $|y| \geq y_0$. In this sense the GRS model can be considered 
as a regularized version of the RS model. 
The fact that the GRS model violates the weakest form of a positive
energy condition seems to indicate that the GRS background spacetime
may be unstable. However, this is not all of the story. Whether or
not the GRS background spacetime is really unstable should be checked
by the perturbation study, which we shall do below.

\section{Linearized perturbations in GRS model}
 
The GRS model with a positive tension domain wall at $z=0$ and 
two negative tension walls at $z=\pm z_0$~\cite{GRS} perpendicular 
to the extra fifth direction can be described by the following 
action:
\beq
I = \int d^4x \int^{\infty}_{-\infty} dz \Big [
    \fr{1}{16\pi G_5} \sr{-\hat{g}} (\hat{R} -2\Lambda )
    -\sr{-\hat{g}_B} \sigma (z) +{\cal L}_M  \Big ] .
\eeq
Although we used the horospherical coordinates $x^M =(x^{\mu},y)$
in the previous section, we introduce here the conformally flat 
coordinates $x^M=(x,z)$ for 
our perturbative analysis. Here $G_5$ is the 5D Newton's constant, 
$\Lambda$ the bulk cosmological constant of five dimensinal 
spacetime, $\hat{g}_B$ the determinant of the metric describing the 
brane, and the tension of the branes $\sigma (z)=\sigma 
\delta^{\rm GRS} (z)$ with $\sigma =3/4\pi G_5L$. $I_M = 
\int d^4xdz {\cal L}_M$ denotes the matter action, and it 
contributes only in the linearized level. 
 
If we introduce a conformal factor as follows
\beq
ds^2 = \hat{g}_{MN} dx^Mdx^N = H^{-2} g_{MN} dx^Mdx^N ,
\eeq
the field equation becomes~\cite{IV,KM}
\beqa
\mbox{} & & G_{MN} +3\fr{\nabla_M\nabla_N H}{H} -3g_{MN}
\Big [ \fr{\nabla_P\nabla^P H}{H} 
-2\fr{\nabla_P H \nabla^P H}{H^2} \Big ] \nonumber   \\
& & = 8\pi G_5 \Big [ -\fr{\Lambda}{8\pi G_5H^2} g_{MN}
-\fr{\sr{-g_B}}{\sr{-g}} \fr{|H|}{H^2} \sigma (z) g_{\mu\nu}
\delta^{\mu}_M \delta^{\nu}_N -\fr{2}{\sr{-\hat{g}}}
\fr{\delta I_M}{\delta \hat{g}^{MN}} \Big ] 
\label{FE0}
\eeqa
with the Einstein tensor $G_{MN}$ constructed from the metric
$g_{MN}$. Now it is straightforward to see that, 
in the absence of matter source except for the domain walls 
themselves (i.e., $\delta I_M/ \delta \hat{g}^{MN} =0$), the 
most general solution having a 4D Poincar\'e symmetry is 
\beq
ds^2 = H^{-2}(z) (\eta_{\mu\nu}dx^{\mu}dx^{\nu} +dz^2),
\eeq
where 
\beq
H(z) = 
\left\{
\begin{array}{ll}
\fr{1}{L}|z|+1, \\  
\fr{1}{L}|z_0|+1,
\end{array}\right.
\qquad
\Lambda (z)=
\left\{
\begin{array}{ll}
-\fr{6}{L^2}  \qquad {\rm for} \qquad |z| \leq z_0, \\
0  \qquad\qquad {\rm for} \qquad |z| \geq z_0 . 
\end{array}\right.
\eeq

Let us consider metric fluctuations around this background 
spacetime as follows~\cite{IV,MKL}:  
\beq
g_{MN} = \eta_{MN} +h_{MN}.
\eeq
Defining $\bar{h}_{MN} =h_{MN}-\fr{1}{2}\eta_{MN} h$ where 
$h=\eta^{MN}h_{MN}$, the linearized perturbation equation of
Eq.~(\ref{FE0}) is  
\beqa
\mbox{} && -\fr{1}{2} \Box \bar{h}_{MN} +\partial_{(M}
\partial^P\bar{h}_{N)P} -\fr{1}{2} \eta_{MN} \partial^P
\partial^Q \bar{h}_{PQ} -\fr{3\partial^PH}{2H}(\partial_M
h_{NP} +\partial_N h_{MP} -\partial_P h_{MN}) \nonumber \\
&& -3\eta_{MN} \Bigg [ \Big (-\fr{\partial^P\partial^QH}{H}  
+2\fr{\partial^PH\partial^QH}{H^2}\Big )h_{PQ} -\fr{\partial^QH}{H} 
\partial^P \bar{h}_{PQ} \Bigg ] 
-3 \Bigg (\fr{\Box H}{H} -2\fr{\partial_PH
\partial^P H}{H^2} \Bigg ) h_{MN}  \nonumber  \\
&& +8\pi G_5 H^{-2} \Bigg \{ \fr{\Lambda (z)}{8\pi G_5} h_{MN} +|H| 
\sigma (z) \Big [ \fr{1}{2}(\eta^{\alpha\beta}h_{\alpha\beta}  
-\eta^{PQ}h_{PQ}) \eta_{\mu\nu}\delta^{\mu}_M\delta^{\nu}_N 
+\delta^{\mu}_M\delta^{\nu}_N h_{\mu\nu}\Big ] \Bigg \} 
\nonumber  \\
&=& 8\pi G_5 T_{MN}, 
\label{LFE0}
\eeqa
where the linearized matter source 
$T_{MN}=-\delta (2\delta I_M /\sr{-\hat{g}} \delta \hat{g}^{MN})$ 
is included, and $\Box =\eta^{MN} \partial_M \partial_N$. 
Taking the 5D harmonic gauge condition,
\beq
\partial^M \bar{h}_{MN} =0 \qquad\qquad {\rm or} \qquad 
\qquad   \partial^M h_{MN} =\fr{1}{2} \partial_N h ,
\label{HG}
\eeq
the linearized equation becomes
\beqa
\mbox{} && \Box h_{MN} +3\fr{\partial_5H}{H} (\partial_M
h_{5N} +\partial_N h_{5M} -\partial_5 h_{MN}) 
+\Big (-\fr{8(\partial_5H)^2}{H^2} +\fr{4\partial^2_5H}{H}
\Big ) h_{55}\eta_{MN} \nonumber  \\
&& -6\fr{\partial^2_5H}{H} \Big [ \fr{1}{2}h_{55}\eta_{\mu\nu}
\delta^{\mu}_M \delta^{\nu}_N -h_{5\mu}(\delta^{\mu}_M\delta^5_N
+\delta^5_M\delta^{\mu}_N) \Big ] = -16\pi G_5 
\Big (T_{MN} -\fr{1}{3}\eta_{MN}T^P_P \Big ) . 
\label{LFE1}
\eeqa
Notice that we do not impose the trace free condition $h=h^M_M=0$.
In components, the above equations become
\beqa
\mbox{} && (\Box +3\fr{\partial_5H}{H} \partial_5 )h_{55}
+\Big (-\fr{8(\partial_5H)^2}{H^2} +\fr{4\partial^2_5H}{H}
\Big ) h_{55} = -\fr{32\pi G_5}{3}\Big (T_{55} -\fr{1}{2}
T^{\rho}_{\rho} \Big ), 
\label{h55}  \\ 
&& (\Box +6\fr{\partial^2_5H}{H}) h_{5\mu} +3\fr{\partial_5H}
{H}\partial_{\mu} h_{55} = -16\pi G_5 T_{5\mu},   
\label{h5m} \\
&& (\Box -3\fr{\partial_5H}{H}\partial_5 )h_{\mu\nu} +3\fr{
\partial_5H}{H}(\partial_{\mu}h_{5\nu} +\partial_{\nu}h
_{5\mu}) +\Big (-\fr{8(\partial_5H)^2}{H^2} 
+\fr{\partial^2_5H}{H}\Big )h_{55}\eta_{\mu\nu} \nonumber   \\
&& = -16\pi G_5 \Big (T_{\mu\nu} 
-\fr{1}{3}\eta_{\mu\nu}T^P_P \Big ) . 
\label{hmn}
\eeqa
Note that $\bar{h}=\eta^{MN}\bar{h}_{MN}=-\fr{3}{2}h$. 
 
For simplicity, we consider a case of $h_{5\mu}=T_{5\mu}=0$.
Then Eq.~(\ref{h5m}) implies that $h_{55}$ is a function of the 
$z$-coordinate only. Thus, one can rescale the $z$-coordinate so
that $h_{55}=0$. In other words, we take the Gaussian normal 
gauge (i.e., $h_{5\mu}=h_{55}=0$) without the trace free condition.
Subsequently, Eq.~(\ref{h55}) shows that it is
neccessary for the matter source to satisfy the following relation 
\beq
T_{55} = \fr{1}{2}T^{\mu}_{\mu}.
\label{tr55}
\eeq
We see that this is exactly the stabilization condition for the 
extra dimension implemented in Refs.~\cite{KKOP,KR}. 
Note that $T^P_P=T^{\mu}_{\mu} +T_{55}= \fr{3}{2}T^{\mu}_{\mu}$.
Then Eq.~(\ref{hmn}) becomes
\beq
(\Box -3\fr{\partial_5H}{H}\partial_5 )h_{\mu\nu} 
= -16\pi G_5 \Big (T_{\mu\nu} -\fr{1}{2}\eta_{\mu\nu}T \Big ), 
\label{hmn2}
\eeq
where $T=T^{\mu}_{\mu}$. Since $h=\eta^{\mu\nu}h_{\mu\nu}
+h_{55}=h^{\mu}_{\mu}$, the harmonic gauge condition 
in Eq.~(\ref{HG}) gives 
\beq
\partial^{\mu} (h_{\mu\nu} -\fr{1}{2}\eta_{\mu\nu}h^{\rho}_{\rho}) 
=0, \qquad\qquad 
\partial_5 h^{\rho}_{\rho} =0.
\label{HG2}
\eeq
Since the trace of Eq.~(\ref{hmn2}) is
\beq
(\Box -3\fr{\partial_5H}{H}\partial_5 )h^{\mu}_{\mu}
= 16\pi G_5 T ,
\label{Trace0}
\eeq
Eq.~(\ref{hmn2}) can also be written as 
\beq
(\Box -3\fr{\partial_5H}{H}\partial_5 )(h_{\mu\nu} -\fr{1}{2}
\eta_{\mu\nu}h^{\rho}_{\rho}) = -16\pi G_5 T_{\mu\nu}. 
\label{hmn3}
\eeq
Thus, the gauge condition in Eq.~(\ref{HG2}) strictly leads to
the source conservation law
\beq
\partial^{\mu}T_{\mu\nu}=0.   
\label{CL}
\eeq
This is a relic of the 4D Poincar\'{e} symmetry in the linearized
level. Note also that, using the second gauge condition 
in Eq.~(\ref{HG2}), we find from Eq.~(\ref{Trace0})
\beq
\Box_4 h^{\mu}_{\mu} = 16\pi G_5 T^{\mu}_{\mu} \qquad
\qquad   {\rm with} \qquad \Box_4 =\eta^{\mu\nu}\partial_{\mu}
\partial_{\nu} .  
\label{Trace}
\eeq
This means that the trace $h$ can propagate on the brane if 
one includes the matter source. Note, however this corresponds to
the massless scalar propagation. Furthermore, by taking 
$\partial_5$ on Eq.~(\ref{Trace}) and using $\partial_5
h^{\mu}_{\mu}=0$, we have additional constraints for the source 
\beq 
\partial_5T^{\mu}_{\mu}=\partial_5 T_{55}=0. 
\eeq
Therefore, for the consistency of linearized equations, we find 
that $T^{\mu}_{\mu}$ and $T_{55}$ of the matter source are to be 
constant in the extra dimension. Our uniform matter of 
$T^{\mu}_{\mu} =T^{\mu}_{\mu}(x)$ is lead to 
keep up the trace $h$ with a physical variable. In the absence 
of the matter source (more precisely, $T^{\mu}_{\mu}=0$), the 
trace $h$ belongs to a gauge degree of freedom and thus it can be 
gauged away. In this case, $h$ is not a physical variable. Further 
our matter source with uniform trace cannot affect the massive KK 
modes because it is independent of $z$. 
  
Defining $h_{\mu\nu}(x,z) = H^{3/2}\psi (z) \hat{h}_{\mu\nu}(x)$, 
$\psi (z)$ satisfies the following Schr\"{o}dinger-like equation
\beq
\big [ -\fr{1}{2}\partial^2_5 +\fr{15(\partial_5H)^2}{8H^2} 
-\fr{3\partial^2_5H}{4H}\big ] \psi (z) 
= \fr{1}{2}m^2_h \psi (z),
\label{gravi}
\eeq
where the seperation constant $m_h$ plays as the mass of the 4D 
gravitational wave $\hat{h}_{\mu\nu}(x)$. 
Then Eq.~(\ref{hmn2}) becomes
\beq
(\Box_4 -m^2_h)h_{\mu\nu}
=-16\pi G_5 (T_{\mu\nu} -\fr{1}{2} \eta_{\mu\nu} T),
\label{hmn4}
\eeq

\section{No tachyonic condition}

Now we are in a position to discuss ``no tachyonic condition." 
Defining $2\ln H(z)=B(z)$, Eq.~(\ref{gravi}), which determines 
the spectrum of KK excitations, can be written as 
\beq
-\fr{d^2 \psi (z)}{dz^2} + [ \fr{9}{16} {B'(z)}^2 -\fr{3}{4}B''(z) ]
\psi (z) = m_h^2 \psi (z).
\eeq
This can be further taken into the factorization 
form~\cite{CEH,DFGK}
\beq
[ -\fr{d}{dz} +\fr{3}{4} B'(z) ] [ \fr{d}{dz} +\fr{3}{4} B'(z) ]
\psi (z) =m_h^2 \psi (z)
\eeq
which has the form of the supersymmetric quantum mechanics
$Q^{\dagger}Q \psi (z) =m_h^2 \psi (z)$, with $Q = \fr{d}{dz}
+\fr{3}{4} B'(z)$.
 
The lowest energy state is the zero-energy state which satisfies
the supersymmetric condition $Q \hat{\psi}_0 (z)=0$,
$\hat{\psi}_0 (z) =e^{-\fr{3}{4}B(z)} = H^{-3/2} (z)$. 
This does not correspond to the normalizable spin-2 propagation.
Hence there is no negative energy graviton modes (tachyon modes)
in the GRS model. We prove that $m^2_h \geq 0$.

\section{No ghost state}

Now we examine the graviton propagator on the positive brane at 
$z=0$ by considering $h_{\mu\nu}(x,0) \sim \hat{h}_{\mu\nu} (x)$ 
only. It requires the bilinear forms of the source with the inverse
propagator to isolate the physical modes. As the present analysis
is on the classical level, we express $\hat{h}_{\mu\nu}$
in terms of source. Taking Fourier transformation for 
Eq.~(\ref{hmn4}) to momentum space results in
\beq
\hat{h}_{\mu\nu} (p) = \fr{16\pi G_5}{p^2+m^2_h} \Bigg [
T_{\mu\nu}(p) -\fr{1}{2}\eta_{\mu\nu} T(p) \Bigg ] .
\eeq
Then the one graviton exchange amplitude for the source
$T_{\mu\nu}$ is given by~\cite{MKL,KR}
\beq
A^{\rm class} = \fr{1}{4} \hat{h}_{\mu\nu}(p)
T^{\mu\nu}(p) = \fr{4\pi G_5} {p^2+m^2_h}
(T^{\mu\nu}T_{\mu\nu} -\fr{1}{2} T^2).
\label{sgea}
\eeq

In order to study the massive states, it is best to use
the rest frame~\cite{SS} in which
\beq
p_1 \neq 0, \qquad\qquad  p_2 =p_3 =p_4 =0.
\label{massivefr}
\eeq
Considering Eqs.~(\ref{CL}) and (\ref{massivefr}) leads to
the following source relations
\beq
T_{11} =T_{12} =T_{13} =T_{14} =0.
\eeq
Thus, one obtains
\beq
T^{\mu\nu}T_{\mu\nu} -\fr{1}{2}T^2 =|T_{+2}|^2 +|T_{-2}|^2 
+|T_{+1}|^2 +|T_{-1}|^2 +T_{44}\big [ \fr{1}{2}T_{44}  
-(T_{22}+T_{33}) \big ],
\label{sgeam} 
\eeq
where the first two terms correspond to the exchange of graviton
with helicity-2 $T_{\pm 2}=\fr{1}{2}(T_{22}-T_{33}) \pm iT_{23}$,
and the third and fourth terms are the exchange of the graviphoton 
with helicity-1 $T_{\pm 1}=T_{24} \pm iT_{34}$. We note here that 
the last term in the above equation is {\it not} positive definite. 
This means that there exist ghost states (negative norm states) 
in general. However, if one requires
\beq
T_{44} =2(T_{22}+T_{33}),
\label{gf}
\eeq
one immediately finds that
\beq
T^{\mu\nu}T_{\mu\nu} -\fr{1}{2}T^2 =|T_{+2}|^2 +|T_{-2}|^2
+|T_{+1}|^2 +|T_{-1}|^2
\eeq
with all positive norm states and without helicity-0 states 
(graviscalars). In the case of $2(T_{22}+T_{33})= aT_{44}$ with
$a < 1$, we find no ghost states, but there exist the graviscalars
which arise from the diagonal elements of $T_{\mu\nu}$.  

In the limit of $m^2_h \rightarrow 0$, the graviphoton propagation
can be decoupled from the brane~\cite{vVZ}. Hence we can neglect 
$|T_{\pm 1}|^2$-terms. Finally the amplitude takes the form
\beq
A^{\rm class}_{m^2_h \rightarrow 0} = \lim_{m^2_h \rightarrow 0}
\fr{4\pi G_5}{p^2_1 +m^2_h}
\big [ |T_{+2}|^2 +|T_{-2}|^2 \big ] ,
\label{sgea0}
\eeq
which corresponds to the massless spin-2 amplitude.

\section{Discussion}

Naively, it is conjectured that, if a certain ${AdS}_5$ spacetime 
does not satisfy the WEC, this may belong to an unstable manifold. 
However, this is not all of the story. We believe that the stability 
analysis of the given spacetime in curved space is based on the 
perturbation study around the background spacetime. This corresponds 
to testing the reliability of its linearized approximation. Hence we 
investigate the stability of the GRS spacetime along this line. 
Here, as in usual Minkowski background, we require two conditions for 
the stable background: no tachyon and no ghost states. Especially, 
we guarantee that the GRS background is stable against the small 
perturbation because it has no tachyon and no ghost states. 
No tachyon condition is easily checked by observing the 
Shr\"{o}dinger equation for the massive KK modes. However, showing 
that there is no ghost states in the GRS model is a non-trivial task. 
This is because, in this model, the massless spin-2 propagation 
(graviton) can be only described by the massive KK modes with 
resonance. 

The main problem is to cancel the unwanted extra polarization 
in the quasi-localization of 4D gravity. This is done by introducing 
both the trace ($h$) and the matter source with uniform trace  
($T^{\mu}_{\mu}$) at the linearized level. In the conventional RS 
approach, the trace ($h$) is just a gauge-dependent scalar and 
hence it can be gauged away. However, including the matter with 
uniform trace, this plays a role of $\xi^5$ in the brane-bending 
model with the localized source~\cite{GT}. This is because 
$h$ ($\xi^5$) satisfy the nearly same massless equations of 
$\Box_4 h = 16\pi G_5 T^{\mu}_{\mu}$ ($\Box_4 \xi^5 = 
\fr{8\pi G_5}{6} S^{\mu}_{\mu}$ in Ref.~\cite{GT}). And the 
comparison of $\bar{h}_{\mu\nu} =h_{\mu\nu}-\fr{1}{2}\eta_{\mu\nu} 
h$ with $\bar{h}_{\mu\nu} =h^{(m)}_{\mu\nu} +2L^{-1}
\eta_{\mu\nu}\xi^5$ in Ref.~\cite{GT} confirms the close 
relationship between $h$ and $\xi^5$. If $T^{\mu}_{\mu}=0$, 
one finds from Eq.~(\ref{sgea}) that the massive spin-2 states have 
5 polarizations with all positive norm states~\cite{MKL}. Here, in 
the case of $h \neq 0$, $T^{\mu}_{\mu} \neq 0$, requiring the 
additional condition $T_{44} =2(T_{22}+T_{33})$ in Eq.~(\ref{gf}), 
we find the massless spin-2 state with 2 polarizations in the
limit of $m^2_h \to 0$. In this case the ghost states disappear. 

In conclusion, it turns out that the GRS spacetime is stable 
against the small perturbation if $T^{\mu}_{\mu} \neq 0$,  
$h \neq 0$ with $T^{\mu}_{\mu}=2T_{55}=3(T_{22}+T_{33})$. 
On the other hand, the RS spacetime is stable under the RS 
gauge ($h=h_{5\mu}=h_{55}=0, \partial_{\mu}h^{\mu\nu}=0$) 
and $T^{\mu}_{\mu}=0$~\cite{MKL}.

\section*{Acknowledgments}

The authors thank Hyungwon Lee for helpful discussions.  
This work was supported by the Brain Korea 21 Programme, Ministry of
Education, Project No. D-0025.

\end{document}